# Modes of wall induced granular crystallisation in vibrational packing

Weijing Dai[1], Joerg Reimann[2], Dorian Hanaor[3], Claudio Ferrero[4], Yixiang Gan[1,*]

[1] School of Civil Engineering, The University of Sydney, NSW 2006, Australia.

[2] Karlsruhe Institute of Technology, P.O. Box 3640, 76021 Karlsruhe, Germany.

[3] Fachgebiet Keramische Werkstoffe, Technische Universität Berlin, Germany.

[4] ESRF-The European Synchrotron, Grenoble, France.

**Abstract**

Granular crystallisation is an important phenomenon whereby ordered packing structures form in granular matter under vibration. However, compared with the well-developed principles of crystallisation at the atomic scale, crystallisation in granular matter remains relatively poorly understood. To investigate this behaviour further and bridge the fields of granular matter and materials science, we simulated mono-disperse spheres confined in cylindrical containers to study their structural dynamics during vibration. By applying adequate vibration, disorder-to-order transitions were induced. Such transitions were characterised at the particle scale through bond orientation order parameters. As a result, emergent crystallisation was indicated by the enhancement of the local order of individual particles and the number of ordered particles. The observed heterogeneous crystallisation was characterised by the evolution of the spatial distributions via coarse-graining the order index. Crystalline regimes epitaxially grew from templates formed near the container walls during vibration, here termed *the wall effect*. By varying the geometrical dimensions of cylindrical containers, the obtained crystallised structures were found to differ at the cylindrical wall zone and the planar bottom wall zone. The formed packing structures were quantitatively compared to X-ray tomography results using again these order parameters. The findings here provide a microscopic perspective for developing laws governing structural dynamics in granular matter.

# Introduction

The behaviour of granular matter subjected to shear, vibration, flow and mixing is of tremendous significance in diverse applications involving handling and processing materials in particulate form [1,2]. The evolution of granular packing is particularly important as system behaviour and effective properties are closely correlated to the packing structure [3]. Recently, granular crystallisation under vibration has drawn increasing interests [4-10]. Similar to its counterpart in materials science, crystallisation in granular matter can be characterised by the formation of ordered structures. Analogous to heat in materials science, vibration acts as an energy source to agitate particles to jump around and form crystalline structures. However, as granular matter is an athermal and generally repulsive system, the underlying mechanisms differ from bonding at the atomic level. For decades, vibration has been utilised to excite granular matter for fluidisation, segregation and packing [11-14], and is thus utilised in this study to explore how granular crystallisation occurs. Early studies show how internal dynamics of particles determine the macroscopic behaviour of entire granular systems [15-17]. Depending on the intensity of vibration, the manifest macroscopic phenomena during vibration are compaction [18-23] and convection [24-27]. However, granular crystallisation lies at the intersection of these two phenomena, as it requires a certain fluidisation to facilitate particle rearrangement, yet results in a compacted state.

Experimentally, granular crystallisation has been studied by vibrating existing packings [7,28] or by adding particles at controlled rates to horizontally or vertically vibrating boxes [4,29]. In the former scenario, a gradual deceleration of agitation, analogous to annealing, plays an important role in the emergent crystallisation, as do pre-set templates and particular container geometries in the latter case [4]. The use of three-dimensional vibration, rather than uniaxial, is found to further enhance crystallisation by disrupting granular arching [22]. The behaviour of sinusoidal-vibration driven rearrangement of granular packing is manipulated by filling rate and vibration intensity $\Gamma$, defined as $\Gamma = A(2\pi f)^2/\text{g}$, where g is the gravitational acceleration, and $A$ and $f$ are the vibration amplitude and frequency, respectively. Maximum crystallisation is found to occur for an intermediate value of $\Gamma$ [22,30,31]. Earlier works have demonstrated the possibility to control granular packing structures through geometry parameters, cohesion and agitation, with the minimisation of internal energy proposed to drive crystallisation. However, a mechanistic understanding of how granular crystallisation occurs and develops under simple and continuous vibration still remains elusive. Insights can be gained from the densification of granular matter, where mechanisms have been discussed in terms of the preservation and reconstruction of contact networks [32] and pore size distribution [18,33].

In order to establish accurate mechanisms, numerical simulations, such as discrete element methods and Monte Carlo methods, are implemented together with X-ray computed tomography (XCT) to examine structural evolution [23] and the appearance of crystalline structures [9,10]. In recent studies, the emergence of crystallisation by vibration was identified for conditions where the random close packing fraction (0.64) was exceeded [34,35]. By examining the internal structure of various packings formed under different vibration protocols, the formation of polytetrahedral patterns and octahedral cavities was shown to impart geometrical frustration [10,36]. Besides, enabled by the use of high-speed cameras, the role of boundaries in the nucleation of crystalline regimes in two-dimensional systems has been studied across multiple scales in terms of pattern formation and densification as well as their relation to grain mobility [37] and energy dissipation [38]. Three-dimensional systems with planar, convex and concave boundaries have been recently investigated experimentally [35,39], which implies the significance of boundary geometries.

Through well-developed frameworks used to describe analogous crystallisation from glasses or gels in the domain of materials science [40-43], the field of granular crystallisation remains ripe for further exploration, with a view towards enhancing high-value packing-dependent properties [34,44]. The present study explores transient states of crystallisation, using a discrete element method (DEM) to simulate dynamic behaviour of vibrated granular matter. Different container geometries were used to examine the boundary influence, with results compared to XCT data. Simulations show that in the absence of cohesion or attraction, granular matter exhibits a clear tendency to crystallise under vibration. Different modes of crystallisation were identified through the analysis on the packing structure transitions. The favoured propagation direction of crystallisation is examined here as a function of container geometry as is the resulting spatial distribution of crystallites.

## Simulation setup

### Discrete element method

Dynamics of granular matter subjected to external vibration is simulated by the open source software LIGGGHTS [45] based on DEM. In this method, the propagation of external agitation is interpreted as the result of inter-particle collisions, causing the motion of granular particles. For an individual collision between particles $i$ and $j$, the normal component $\boldsymbol{F}_n^{i,j}$ and tangential component $\boldsymbol{F}_t^{i,j}$ of the inter-particle force are described by the following equations,

$$\boldsymbol{F}_n^{i,j} = k_n^{i,j}\,\boldsymbol{\delta}_n^{i,j} - \gamma_n^{i,j}\,\boldsymbol{v}_n^{i,j},\; \boldsymbol{F}_t^{i,j} = k_t^{i,j}\,\boldsymbol{\delta}_t^{i,j} - \gamma_t^{i,j}\,\boldsymbol{v}_t^{i,j}, \tag{1}$$

where $k_n^{i,j}$ and $\gamma_n^{i,j}$ and ($k_t^{i,j}$ and $\gamma_t^{i,j}$) are contact stiffness and viscoelastic damping coefficient for normal and (tangential) contact, respectively. These quantities are derived from the Hertz-Mindlin contact theory, and thus depend on the instantaneous contact configuration. Here, $\boldsymbol{\delta}_n^{i,j}$ is the overlap distance, $\boldsymbol{v}_n^{i,j}$ and ($\boldsymbol{v}_t^{i,j}$) are the relative velocities in the normal and tangential directions, and $\boldsymbol{\delta}_t^{i,j}$ is the tangential displacement vector between particles $i$ and $j$. In addition, $\boldsymbol{F}_t^{i,j}$ is limited by the Coulomb friction limit, $|\boldsymbol{F}_t^{i,j}| \leq \mu |\boldsymbol{F}_n^{i,j}|$, in which $\mu$ is the friction coefficient.

To investigate the influence of container boundaries on granular crystallisation, mono-dispersed frictionless spheres were generated in cylindrical containers of different radii and heights. Then, the evolution of the packing structure inside these granular media during continuous sinusoidal vibration was simulated and characterised to study the crystallisation process. The simulations followed a similar scheme. Firstly, 5000 particles of size $d$ were randomly dispersed in a finite cylindrical container with a diameter $D$ and were allowed to settle under gravity. This state was considered as the initial state before vibration. Sinusoidal vibration function with an amplitude $A$ and frequency $f$ was introduced by moving the bottom of the cylindrical container in the axial direction along the gravitational direction. At the upper surface of the granular matter, a lid that could move freely in axial direction confined the granular matter in a negligible pressure of a few Pa. This lid is particularly used to flatten the upper surface for structural characterisation, e.g. packing fraction and Voronoi tessellation. The lid was only pushed upwards at the onset of the vibration but later remained in contact with the top layer of particles. The vibration was applied over 1000 periods ($T = 1/f$), followed by sufficient relaxation to reach equilibrium. The vibration amplitude was varied to study the influence of the energy input on the crystallisation. Table 1 summarises the material and simulation parameters used in the study. The diameters and the initial heights of the cylindrical volumes were varied to study the geometrical influence on the crystallisation process. $D/H$=30/75 (with $D/d$=13, $H/d$=33) is characteristic for a rather slender container, while the case $D/H$=60/19 (with $D/d$=26, $H/d$=8) represents a rather flat one. The chosen geometrical parameters $D/H$=30/75 and $D/H$=50/25 are similar to those investigated by XCT [39] for the purpose of reliable comparison between simulation and experiment.

Table 1 Parameters for DEM simulations

| Parameter | Value |
|---|---|
| Young's modulus, $E$ | 63 (GPa) |
| Poisson ratio, $\nu$ | 0.2 (-) |
| Density, $\rho$ | 2230 (kg/m$^3$) |
| Friction coefficient, $\mu$ | 0.0, 0.2, 0.5 (-) |
| Coefficient of restitution | 0.6 (-) |
| Diameter of sphere, $d$ | 2.3 (mm) |
| Container Diameter/Height, $D/H$ | 30/75, 40/60, 50/25, 60/19 (mm/mm) denoted as D30, D40, D50, D60 |
| Vibration amplitude, $A$ | 0.23 (0.1$d$), 0.46 (0.2$d$) (mm) |
| Vibration frequency, $f$ | 50 (Hz) |
| Gravitational acceleration, g | 9.8 (m/s$^2$) |
| Vibration intensity, $\Gamma$ | 2.3, 4.6 (-) |

**Order characterisation**

In the present study, we used void fraction distributions, coordination numbers, contact angle and radial density distributions as well as Voronoi tessellation to characterise the packings [32,39,46]. However, we concentrate here on the use of bond orientation order parameters to distinguish crystalline structures and represent the transitions between ordered and disordered states. An advantage of this measure is its insensitivity to particle separation, enabling its use for the transient characterisation of moving granular matter. Together, static and dynamic measures are used to describe the structural changes across order transitions.

*Bond orientation order parameters*, initially defined by Steinhardt *et al* [47], represent the rotational symmetry of sphere assemblies as,

$$Q_{lm}(\vec{r}) \equiv Y_{lm}(\theta(\vec{r}), \varphi(\vec{r})), \tag{2}$$

where a bond $\vec{r}$ is defined as a vector that points from the centroid of a central particle to one of its neighbour particles, $Y_{lm}(\theta, \varphi)$ are spherical harmonics, $\theta(\vec{r})$ and $\varphi(\vec{r})$ are the polar and azimuthal angles of the bond in a reference spherical coordinates system, and $l$ and $m$ are integers indicating the order of spherical harmonics with the condition that $l \geq 0$ and $|m| \leq l$. By averaging $Q_{lm}(\vec{r})$ over the $n_b^i$ closest neighbours of a central particle $i$, the following expression is obtained,

$$\hat{q}_{lm}(i) = \frac{1}{n_b^i} \sum_{k=1}^{n_b^i} Q_{lm}^{i,k}(k). \tag{3}$$

In the current study, the number of neighbours $n_b^i$ of a central particle $i$ is selected as 12 [41], the largest coordination number of non-overlapping mono-sized particles, because it gives significantly different feature values between crystalline structures like BCC, HCP, FCC and even icosahedral. Finally, the local bond orientation order for particle $i$ is constructed as [48]

$$Q_l^{\text{local}}(i) \equiv \left( \frac{4\pi}{2l+1} \sum_{m=-l}^{l} |\hat{q}_{lm}(i)|^2 \right)^{1/2}, \tag{4}$$

and

$$\widehat{W}_l(i) = \frac{\sum_{m_1, m_2, m_3} \begin{pmatrix} l & l & l \\ m_1 & m_2 & m_3 \end{pmatrix} \hat{q}_{lm_1}(i) \, \hat{q}_{lm_2}(i) \, \hat{q}_{lm_3}(i)}{\left[ \sum_{m=-6}^{6} |\hat{q}_{lm}(i)|^2 \right]^{3/2}}, \tag{5}$$

where the term in the parentheses is Wigner-3j symbol. $l = 4$ and 6 are widely used due to their unambiguous value for regular structures [41,48].

*Order indices* of neighbourhood configuration of a central particle $i$ are defined on the basis of a vector $\vec{q}_6(i) = [\hat{q}_{6m}(i)]$, with $m =$ -6, -5,…0,…5, 6. The cosine similarity of a pair of such vectors of neighbouring particles $i$ and $j$ is calculated as

$$\text{CosSimi}(i, j) = \mathbf{Re} \left[ \frac{\vec{q}_6(i)}{|\vec{q}_6(i)|} \cdot \frac{\vec{q}_6(j)}{|\vec{q}_6(j)|} \right] = \mathbf{Re} \left[ \frac{\sum_{m=-6}^{6} \hat{q}_{6m}(i) \cdot \hat{q}_{6m}^*(j)}{|\vec{q}_6(i)||\vec{q}_6(j)|} \right] \tag{6}$$

$\text{CosSimi}(i, j)$ between particles $i$ and $j$ is positively correlated with the similarity of their individual neighbourhood configurations, and a pair of connected particles has $\text{CosSimi}(i, j) \geq 0.7$ [49]. Hence, the order of individual particles can be positively represented by the parameter [41],

$$S_6^i = \sum_j^{n_b^i} \text{CosSimi}(i,j), \tag{7}$$

ranging between 0 and 12. Additionally, to describe the overall structural order of an assembly of particles, we make use of [44],

$$F_6 = \frac{1}{N_P} \sum_{i=1}^{N_P} f_6(i), \tag{8}$$

where

$$f_6(i) = \frac{1}{n_b^i} \sum_{j=1}^{n_b^i} \Theta \left[ \text{CosSimi}(i,j) - 0.7 \right], \tag{9}$$

with the step function $\Theta(\cdot)$. Spanning the range 0 to 1, the term $F_6$ is used to characterise the structural evolution during vibration, as an alternative to the conventional packing fraction.

A bottom-up *Coarse-graining* methodology is applied in this work to convert the discrete data set of the order index $S_6^i$ into a higher-scale continuum form to reveal the spatial distribution of order in the structure. In general, the coarse-graining approach used here takes a set of discrete points $\mathbf{P}_i = (x_i, y_i, z_i)$ and their corresponding scalar data $h_i$ as input [50]. Instead of describing the density at any point $\mathbf{P} = (x, y, z)$ by $\rho^{\text{dis}}(\mathbf{P}) = \sum_i h_i \delta(\mathbf{P} - \mathbf{P}_i)$ where $\delta(\Delta \mathbf{p})$ is the Dirac delta function, the coarse-graining approach transforms this discrete field into a continuous one by replacing $\delta(\Delta \mathbf{p})$ with a positive semi-definite function $\varphi(\Delta \mathbf{p})$. This function fulfils the requirement that the integral of the continuous density function

$$\rho^{\text{con}}(\mathbf{P}) = \sum_i h_i \varphi(\mathbf{P} - \mathbf{P}_i), \tag{10}$$

is equal to the sum of $h_i$ for a given volume. The exact form of $\varphi(\Delta \mathbf{p})$ is not determinative but the width $w$ at which $\varphi(\Delta \mathbf{p})$ vanishes holds significance [50]. Here, a three-dimensional Gaussian function is employed,

$$\varphi(\mathbf{P} - \mathbf{P}_i) = \frac{e^{-\frac{1}{2}\left[\left(\frac{x-x_i}{r}\right)^2 + \left(\frac{y-y_i}{r}\right)^2 + \left(\frac{z-z_i}{r}\right)^2\right]}}{(2\pi)^{\frac{3}{2}}}. \tag{11}$$

This function vanishes at $w = 3r$ where $r$ is the radius of particles. This selection of $w$ results from the fact that the order index $S_6^i$ is based on particle ensembles roughly occupying a spherical region with the radius of $3r$.

**Dynamic characterisation**

*Granular-temperature* is often used as a term, analogous to thermal energy at the atomic scale, to describe the kinematics of granular matter. While crystallisation and ordering at the atomic scale is studied with respect to thermal conditions, here we examine the local and global ordering of packing with respect to the dynamic status of particles. Granular temperature quantifies the velocity fluctuation of particles in granular matter rather than being a measure of thermal energy [51]. To obtain the granular temperature, granular matter is first discretised into particle ensembles [52]. For each ensemble of $n$ particles, the average velocity is calculated in $x$, $y$, and $z$ axis as $\bar{v}_{x,y,z} = \frac{1}{n}\sum_{i=1}^{n} v^i_{x,y,z}$, respectively. Then the velocity fluctuation along $x$, $y$, and $z$ axes are derived separately by

$$\text{GT}_{x,y,z} = \frac{1}{n}\sum_{i=1}^{n}(v^i_{x,y,z} - \bar{v}_{x,y,z})^2. \tag{12}$$

Finally, the granular temperature is the mean of the three axial granular temperatures $\text{GT} = \frac{1}{3}(\text{GT}_x + \text{GT}_y + \text{GT}_z)$. To maintain consistency with the structural order characterisation, the ensembles used to derive the granular temperature are chosen as the 12-neighbour configuration used in the $S_6$ calculation. In this way, every particle has its granular temperature as well as its order index $S_6$, and relations between granular temperature and granular crystallisation can be investigated.

## Results

**Transient behaviour of crystallisation**

Granular matter is characterised by packing fraction ($\gamma$) that gives the volume fraction of the particles. However, the packing fraction can only characterise the densification aspect of crystallisation but not the formation of order [43]. Therefore, the structural index $F_6$ [44] is utilised to quantify the overall progress of the order formation during the granular crystallisation.

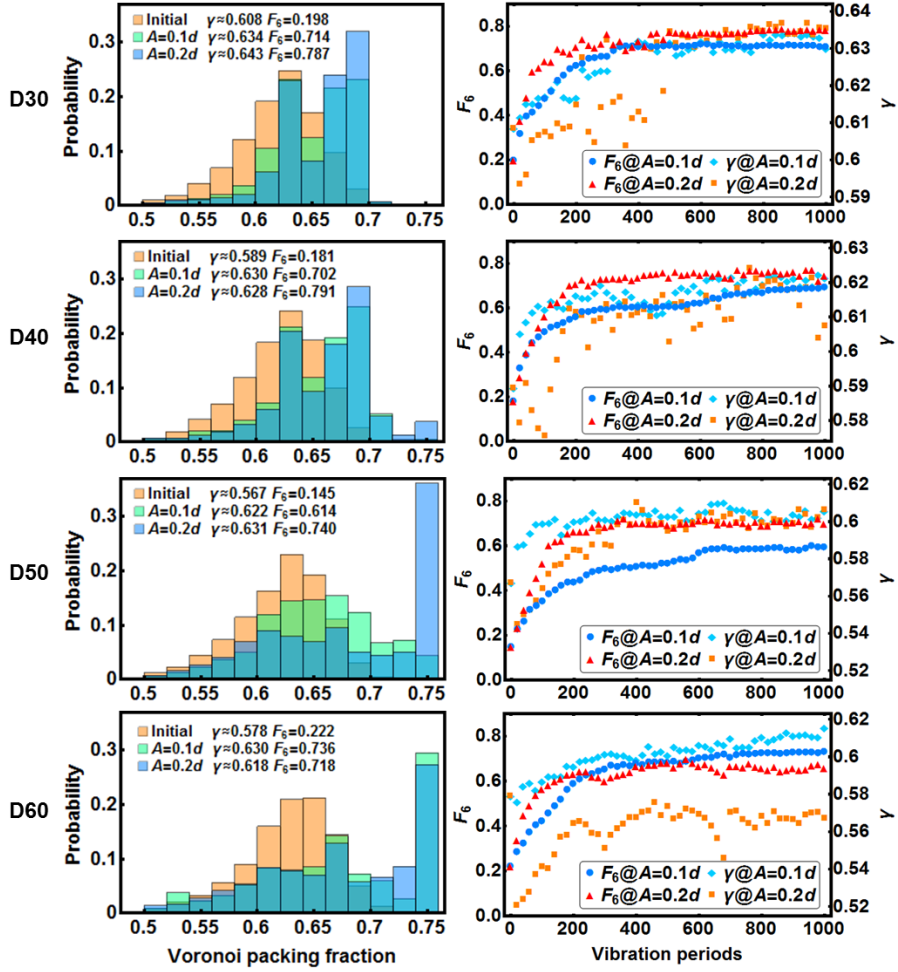

Figure 1. Overall evolution of each granular medium subjected to vibrations of different amplitude. The histograms (colouring with transparency) in the left column show distributions of the Voronoi cell packing fraction of the initial state and two final relaxed states after vibration ($A=0.1d$ and $A=0.2d$) with legends giving the corresponding overall packing fraction $\gamma$ and structural index $F_6$. The corresponding right column plots demonstrate the time variation of the overall at transient states during vibration.

The static and transient states of the granular media are shown in the left and right columns of Figure 1, respectively. In the histograms on left-hand side, the distributions of the Voronoi cell packing fraction are presented with the legends indicating $\gamma$ and $F_6$. Both $\gamma$ and $F_6$ increase in the final relaxed states after vibration, showing that the granular media in the final state become denser and more ordered than the initial state. The larger amplitude leads to a bigger increment in $\gamma$ and $F_6$ except for the case of D60. This exception can be attributed to the strong fluidisation because of the least gravitational potential to overcome, i.e., the least H. These results are in agreement with previously reported DEM studies [46] and experiments [17,39] in which the packing fraction is maximised at an intermediate vibration intensity. Although this is not the aim of the current study, it still can be concluded the maximum $F_6$ can be introduced by an intermediate vibration intensity. While, the packing fraction is only moderately affected, a considerable increase in $F_6$ suggests that the structure of the granular media indeed changes remarkably as a result of vibration. In order to reveal the change in structure between the initial and final states, the Voronoi cell packing fraction was calculated [53]. The distributions are of bell-shape for all the initial states and are characterised by several peaks at the final states. With increasing $D$ and decreasing $H$, a peak at ≈ 0.74 develops, which proves the emergence of a highly crystallised structure.

Before further analysing the crystallised structure, it is of significance to discuss the relation between the two main aspects of granular crystallisation, densification and order formation. The relation has been debated by researchers studying crystallisation in other systems. Whether the densification precedes the formation of order or the other way around is the topic of some discussions [43], with some reports suggesting that these two processes occurred simultaneously [54]. In spite of a limit of established theoretical framework around granular crystallisation, we can address the relation between densification and order formation in granular systems based on the temporal evolution of $\gamma$ and $F_6$ plotted on the right-hand side of Figure 1. The packing fraction $\gamma$ exhibits fluctuations and disruptions with small and large amplitudes, respectively; however, the increasing trend of $F_6$ is invariant. This results are in accordance with the densification mechanisms identified in weak and strong vibration, where the respective push filing and jump filling [32] are both accompanied by particle rearrangement. This decoupled behaviour of $\gamma$ and $F_6$ combined with the purely repulsive interaction between particles suggests that order formation is an independent process promoting granular crystallisation. Since the evolution of packing fraction in vibration has been widely studied [23,39], we focus on order formation in transient states to elucidate the *in-situ* granular crystallisation. Highly ordered structures of elevated $F_6$, implying structural symmetry, lead to symmetrical collisions between particles and uniform dissipation of energy in the granular media, maintaining the systems' stability. Within the vibration duration in this study two stages could be distinguished according to the evolution of $F_6$. The first stage is characterised by the monotonic and rapid increase of $F_6$, which corresponds to instable and disordered structure undergoing an ordering process. The second stage can be treated as a quasi-equilibrium state. The structure is nearly stable because of the mild change of $F_6$. Therefore, the following discussions focus on the interpretation of the order formation during crystallisation process.

**Selective crystallisation by wall effect and epitaxial growth**

To elucidate the microscale mechanism of the granular crystallisation corresponding to the evolution of $F_6$ at the system scale, decomposition of these granular media was performed to investigate the structural transition of individual particles. By employing DEM simulations to track transient states during vibration, the dynamic development of crystalline structures is unveiled, which supplies information difficult to be obtained by XCT. For each particle, its crystalline perfection is characterised by the local structural index $S_6$. High $S_6$ means that particles in an ensemble have similar neighbourhood configurations. In other words, periodic crystalline structures propagate in the high $S_6$ regimes. In order to spatially expose the granular crystallisation process during vibration, $S_6$ of individual particle is fed into the coarse-graining approach to construct spatial density mapping. In the sequential $S_6$ mappings of the granular media experiencing vibration, the dynamic process of crystallisation is reconstructed by the morphology and intensity evolution as shown in Figure 2.

The general and common phenomenon appearing in all cases is that the region of high $S_6$ (shown in red colour) in the density mappings expands and intensifies with vibration, and eventually reaches a quasi-equilibrium state in which only minor changes of ordering could be observed (row by row in Figure 2). Such phenomenon corresponds to the universal $F_6$ evolution discussed before, a rapid increase followed by an asymptote. Similarly, a larger amplitude accelerates the crystallisation rates and produces larger crystalline regimes. In addition, increasing D results in regions of higher $S_6$ value, indicating finer crystalline regimes. This phenomenon can be explained by the structural difference in those crystalline regimes which will be addressed in later sections. Within the scope of current study, the granular crystallisation is initiated automatically once the vibration commences. It could be inferred that highly disordered granular matter is vulnerable to dynamic perturbations like vibration, while the crystallised granular matter at the quasi-equilibrium state manifests the stability of crystalline structure.

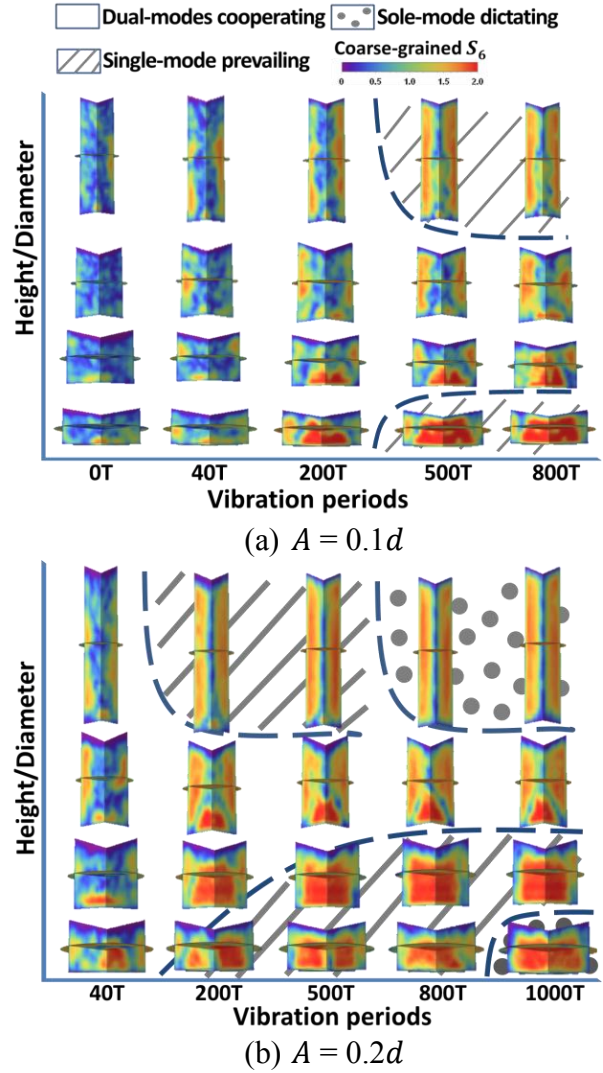

(a) $A = 0.1d$

(b) $A = 0.2d$

Figure 2. Evolutional phase diagram of the crystallisation in granular media of different height-to-diameter ratio. Each $S_6$ density mapping consists of central plane slices of three axes, the colour indicated by coarse grained $S_6$ suggests disorder in violet direction and crystallisation in reddish direction. In the phase diagrams three phases are marked by background filling, (1) dual-modes cooperating; (2) single-mode prevailing; and (3) sole-mode dictating which are approximately determined by the competition between cylindrical and bottom modes.

Regarding the growth of crystalline regimes, a wall effect is identified in all geometries. According to the $S_6$ density mappings at the early periods, reddish regions indicating highly ordered structure always preferentially appear at the bottom walls and cylindrical walls (the second column in Figure 2 (a) and the first column in Figure 2 (b)). This preferential crystallisation can be explained by the existence of a partially ordered layer of particles adjacent to the walls before vibration, which can be seen from the first column in Figure 2 (a). These ordered layers emerge during the settlement of granular matter because the particles need to rest in positions with as many contacts as possible to support themselves under frictionless conditions. After the similar prior crystallisation at walls, the emerging crystallisation becomes dependent on the container boundary configuration, i.e., $D/d$ and $H/d$. Depending on the origin of the crystalline regimes, two crystallisation modes can be differentiated in this period, the cylindrical mode and the bottom mode. The cylindrical mode induces crystalline regimes in a radially inward direction, whereas crystalline regimes grow upwards along the axial direction in the bottom mode. Thus, a competing mechanism between these two modes is introduced for the subsequent crystalline growth stage, which is determined by the H/D ratio. As the H/d decreases and $D/d$ increases, the crystallisation at the bottom wall becomes the favoured mode. Conversely, the cylindrical mode dominates the crystallisation in geometries with small $D/d$ and large $H/d$.

Evolutional phase diagrams were constructed by comparing the $S_6$ density maps of different granular media at the specific vibration duration. Two operational phases exist in the phase diagram of small amplitude ($A=0.1d$) while for the large amplitude ($A=0.2d$) three phases are observed. In the early periods, all granular media exhibit a "dual-mode cooperating" phase where crystallisation progresses on both cylindrical and bottom walls. When the granular media continue to be vibrated, either the cylindrical mode or the bottom mode prevails in crystallisation ("single-mode prevailing" phase), although different crystalline regimes grown in the other modes can still be identified. The exception D40 is always in the dual-mode phase, in which disordered regions act as crystal boundaries and partition different crystalline regimes. The packing structures of the boundary regions are instable and vary with vibration because the mismatch between crystalline regimes grown in two different modes makes it difficult to reach a final state.

Most interestingly, a third operational phase is present for D30 and D60 in the final periods of large amplitude vibration. This phase represents the extreme scenarios where one of the crystallising modes is eliminated, named as "sole-mode dictating" phase. In experiments with optically transparent cylinders, this third operational phase was frequently observed in vibrated particle geometries with large $D/d$ and small $H/d$ by one of the authors (unpublished data). The wall layer at the cylindrical wall was fairly hexagonally packed in dense state for the longest part of the total vibration period. Then, this layer became significantly unstable while at the particle bed surface well defined hexagonal patterns appeared. The final state of the cylinder wall layer was much less regular which is quantified in Experiment $D$ in [39] by the fact that the wall void fraction minimum is the largest of all vibration experiments.

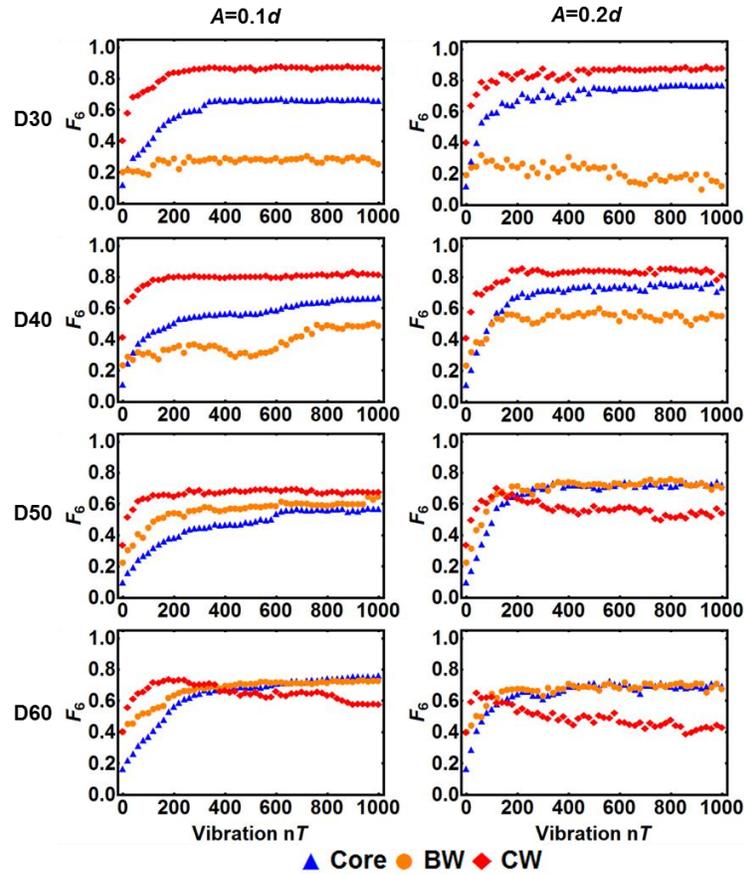

Figure 3 Evolution of $F_6$ for particles groups separated by position. CW – first layer near the cylindrical wall, BW – first layer near the bottom wall and Core – the bulk particles.

Figure 3 shows the variation of $F_6$ in the cylindrical wall layer, the bottom layer and the core region separately with extended vibration periods. It is clearly shown that both wall layers possess higher $F_6$ than the core region at the initial state, proving the existence of partially ordered regimes. Not only for the initial state, has this comparison underlined the significance of the wall effect during the granular crystallisation. The $F_6$ evolution of the two wall layers conforms the phase separation in the evolution diagrams. In dual-mode cooperating phase, the contrast between the patterns is small and both parts exhibit a common trend of increase-to-stabilise behaviour. As the contrast extends or a crossover arises, the granular media move into the single-mode prevailing phase. In regard to the sole-mode dictating phase, it is foreshadowed by the continuous decrease of the $F_6$ in one of the two parts. The decrease of the $F_6$ proves those wall layers become disordered by vibration, wiping out the corresponding mode.

Since very large granular systems (particle number $\gg 10^5$) are not the aims of the current study, we conclude that the granular crystallisation in finite size containers is initiated by the wall effect and progresses in modes depending on the wall geometry. The wall effect of both cylindrical and bottom modes generates quasi two-dimensionally ordered layers with as many contacts as possible which we will subsequently prove to be hexagonal packings. The following processes are the expansion of the hexagonal packing across the wall plane and the epitaxial growth of an adjacent layer. The repetition of these stepwise processes establishes the granular crystallisation in these confined systems. Because of imperfections and the competition with other crystallised layers, each epitaxial layer is smaller than the preceding one, resulting in conical crystalline regimes. From an energetic perspective, the underlying microscale mechanism is proposed as the selection of positions with high number of contacts to enhance the propagation of kinetic energy. Owing to the maximised contacts, the kinetic energy induced by the vibration on the one hand dissipates quickly from one particle to its surroundings in such structures, leaving the particle in a less perturbed state and making the crystalline regimes robust; on the other hand, this kinetic energy is efficiently transferred throughout the crystalline arrangements, triggering the relocation of particles from the disordered regions into particular crystallising positions nearby. Such steps build up a positive feedback loop to promote crystallisation. Therefore, a large amplitude activates more particles near ordered layers and enhances the possibility for these particles to lodge in an energetically favourable position. However, too much energy will reverse those steps, cause drifting particles and deteriorate the crystallisation. Self-nucleation is not the prevailing mechanism due to the repulsive nature of granular matter while the merging of crystalline regimes is commonly observed.

**Granular temperature and crystallisation**

Granular crystallisation can be regarded as the development of structural order. Analogous to structural order, as the uniformity of the velocities of particles in an ensemble increases, we consider it as the development of dynamic order. In this way, the granular temperature is able to quantitatively describe this dynamic order.

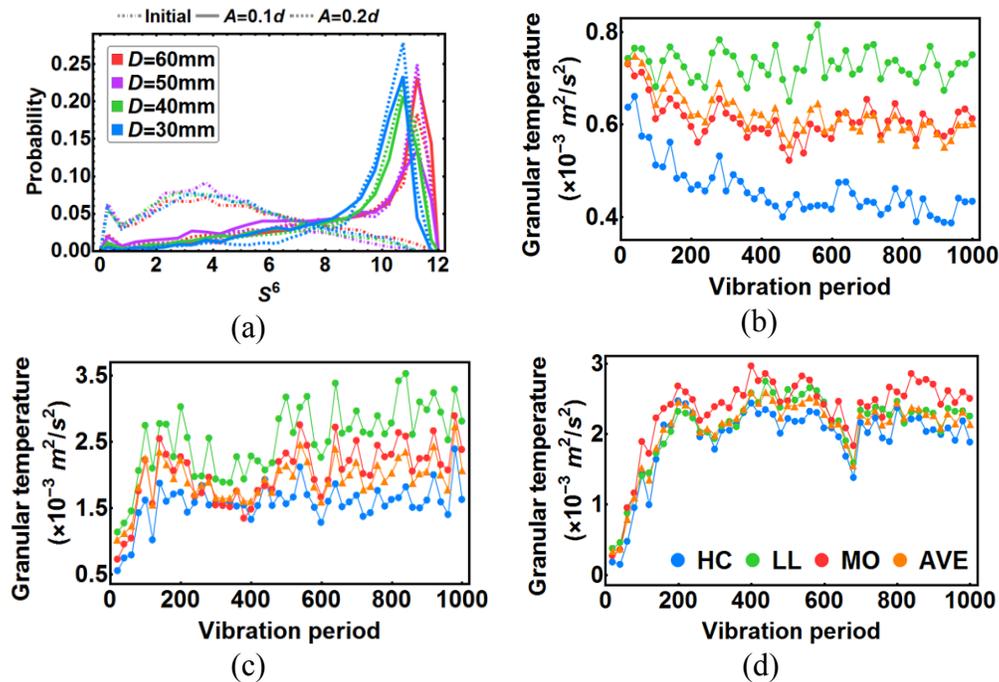

Figure 4 (a) A similar trend in the $S_6$ evolution is observed in all cases. The $S_6$ accumulates between 10 and 12 in the final state, while the peak right shifts as D increases. (b) Typical granular temperature evolution (D50) in the small amplitude vibration scenario. (c) Typical granular temperature evolution (D40) in the large amplitude vibration scenario. (d) Granular temperature evolution in the relatively strong fluidisation case (D60). (b) and (c) have the same legend shown in (d) with HC – highly crystallised, LL – liquid like, MO – moderately ordered, AVE – averaged.

In the temporal evolution of the $S_6$ distributions, a general trend is evident in which a singular peak gradually intensifies during vibration, shown in Figure 4(a). Therefore, we make use of the $S_6$ distributions to examine how the granular temperature evolves under different conditions. By using the particle connection criteria based on the magnitude of $S_6$, the particles are categorised into two groups, the solid-like and liquid-like. Within the solid-like group, the particles are categorised again into two sub-groups, the highly crystallised and moderately ordered, by setting a threshold for highly crystallised particles, i.e., $S_6>10$; because no liquid-like particle possesses $S_6$ which exceeds 10. After such categorisation, the average granular temperature can be extracted for three groups. As discussed in the previous sections, the particles in the boundary regions are supposed to be more active than those in the crystallised regimes. Thus, the dynamic order should be suppressed in the boundary regions of low order. A general observation is that the highly crystallised particles have the lowest average granular temperature but the liquid-like particles have the highest average granular temperature. The distinction of D60 vibrated via large oscillation shown in Figure 4 demonstrates that the dynamic order is suppressed in the over-fluidised state. Thus, we argue that the granular temperature has a transition from divergence to convergence depending on the geometry and the vibration intensity, and further research is required to fully demonstrate it.

This phenomenon is consistent with the discussion on the relation between the propagation of kinetic energy and the order formation in the crystallisation presented in the previous section. Along with $S_6$ rise, the order formation and the particles crystallisation, the local structure of individual particles in granular media becomes periodic and symmetric. As a result, the particle collisions inside such crystalline structures turn out to be counterpoised, causing the structures to attain the ability to suppress velocities deviating from the vibration axis and preventing the particles from over-acceleration. In conclusion, the structural order is the foundation for the dynamic order. The granular crystallisation can stabilise the granular media subjected to agitation by resetting the elevation of the granular temperature.

**Structural characterisation and evolution**

It was demonstrated in the previous sections that the wall effect and the dual crystallisation modes are the major events during vibration. However, $S_6$ can capture the degree of order of the particles but cannot distinguish the structural types of crystalline regimes. Thus, the local bond orientation parameters $Q_6^{local}$, $W_4^{local}$ and $W_6^{local}$ are utilised to differentiate the structure of those regimes. Firstly, $W_6^{local}$ is used to examine whether the cubic-based structures exists based on its sign (+/-) contrast, positive for BCC and negative for HCP and FCC. Nearly all particles in these granular media have negative $W_6^{local}$, denoting that the majority of particles sit in hexagon-based structures (not shown explicitly in the figures). Next, the ($Q_6^{local}$, $W_4^{local}$) coordinates, exhibiting specific coordinates for HCP (0.485, 0.134) and FCC (0.575, -0.159), are employed to characterise the particular structures in the granular media.

In Figure 5, two series of the probability density distributions of ($Q_6^{local}$, $W_4^{local}$) for D30 and D60 demonstrate the typical development pathways of the crystalline structures. The similar broad distribution of the pairs of ($Q_6^{local}$, $W_4^{local}$) indicates the disordered nature of those granular media at the initial state. Three structuring paths are clearly identified in the granular crystallisation. Once the vibration starts, the path leading to a non-typical structure coordinates, neither HCP nor FCC, appears first. Such coordinates represents the hexagonally packed surface particles in finite HCP or FCC structures, calculated from the neighbour configuration lacked half space due to the boundary conditions. In accordance with the previous discussion, this path reveals the tendency towards hexagonal units in the wall layer during vibration. More importantly, its precedence over the other paths proves the priority of the wall effect. When the vibration continues, the HCP and FCC become the main structures emerging during crystallisation. As stated before, the dominating crystallisation mode in the granular media of D30 and D60 is the cylindrical mode and the bottom mode, respectively. According to the density contrast in the HCP and FCC paths between D30 and D60, the cylindrical mode shows a preferential selection of the HCP structure but the bottom mode has little bias on the two paths.

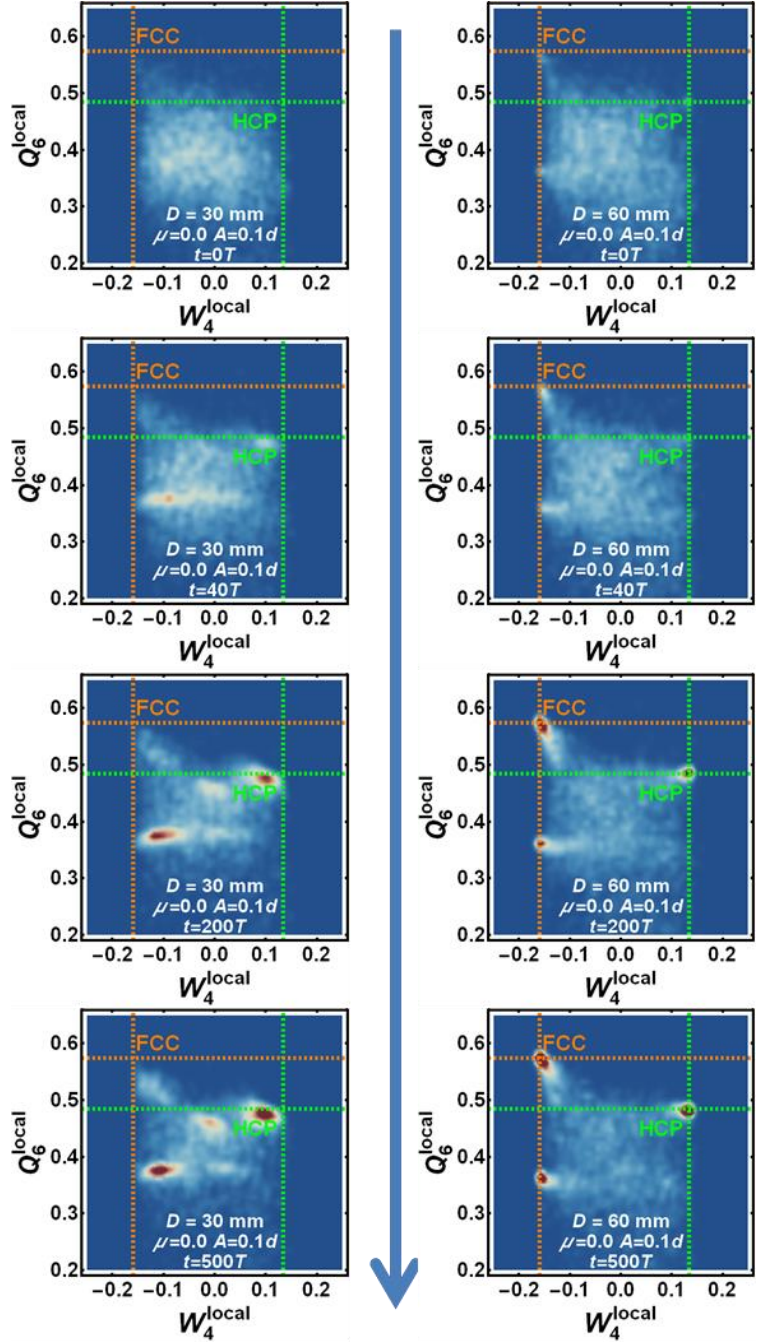

Figure 5. Smoothed probability density histograms of the crystallised structures appearing in the granular media during vibration for D30 and D60. The ($W_4^{\text{local}}$, $Q_6^{\text{local}}$) coordinates are used to characterise the structure types. The intersects of pairs of dashed lines in orange and green are the coordinates of the FCC and HCP, respectively.

The unbiasing in the bottom mode can be explained by the identical two-dimensional hexagonal units in the close packed layers of the HCP and FCC structures but the reason for the incapability of producing FCC structure in the cylindrical mode of large wall curvature is elusive. However, a recent study reported that the mechanical stability of FCC is stronger than HCP [55], resulting in higher resistance to deform in FCC. On the contrary, lower resistance should bring about potential adequateness in HCP to arrange particles in a distorted field like the cylindrical wall without a complete loss of the structural characteristics. Thus, this stronger deformability contributes to the prevalence of HCP in the cylindrical mode. The distinctive crystalline regime in the cylindrical mode is characterised by a deviation from the perfect HCP coordinates introduced by the structural distortion. This cylindrical curvature effect, discussed also in [39], changes the structure greatly from the perfect HCP structure, causing a decrease of the Voronoi cell packing fraction and the coordination number. Nevertheless, as demonstrated in Figure 6, the detailed examination of the 12-neighbour configuration of the distorted particles reveals that the geometrical symmetry is partly maintained in a HCP fashion. HCP segments (blue particles in Figure 6) and rupture particles (yellow particles in Figure 6) are distinguished by the $W_4^{local}$ of rupture particles being close to 0. A layer shift occurs in the neighbour configurations of the segment particles but incomplete layers are the typical feature of the rupture particles. Besides, this distortion causes the weaker ordering of the crystalline regimes near the cylindrical walls when compared with the regimes near the bottom walls shown in Figure 2.

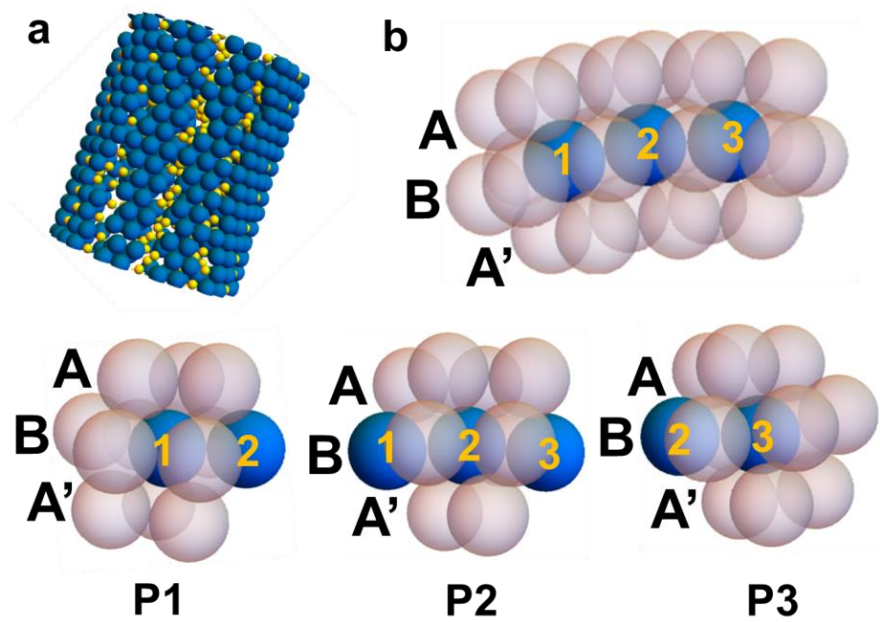

Figure 6 Rupture in the HCP structure near the cylindrical wall in D30. (a) Blue particles are distorted HCP particles ( $0.465 \leq Q_6^{local} \leq 0.505$, $W_4^{local} \geq 0.08$ ) while yellow particles are rupture particles ( $0.465 \leq Q_6^{local} \leq 0.505$, $|W_4^{local}| \leq 0.02$ ) with size scaled by 0.5 for visibility. (b) Typical rupture section with Particle 2 being the rupture particle. P1, P2, P3 are the 12-neighbour configurations of Particle 1, 2, 3, respectively.

With increasing $D$ and decreasing $H$, a stronger tendency towards FCC packing is seen and both HCP and FCC regions exhibit fewer defects. This phenomenon results from the suppression of the cylindrical curvature effect as well as the cylindrical mode, leading to less distorted HCP structure. This can be evidenced by the appearance of more FCC particles (blue) in the cylindrical crystalline regimes with increasing $D$ in Figure 7. Even with the influence of the cylindrical wall, the crystallisation paths show that the tendency for maximising the number of inter-particle contacts still plays a dominant role in granular crystallisation. In the scope of the current study, vibration intensity has a negligible influence on the structuring paths but it increases the efficiency of granular crystallisation according to the quick development of structuring paths and increased density in the ($W_4^{local}$, $Q_6^{local}$) distribution plots.

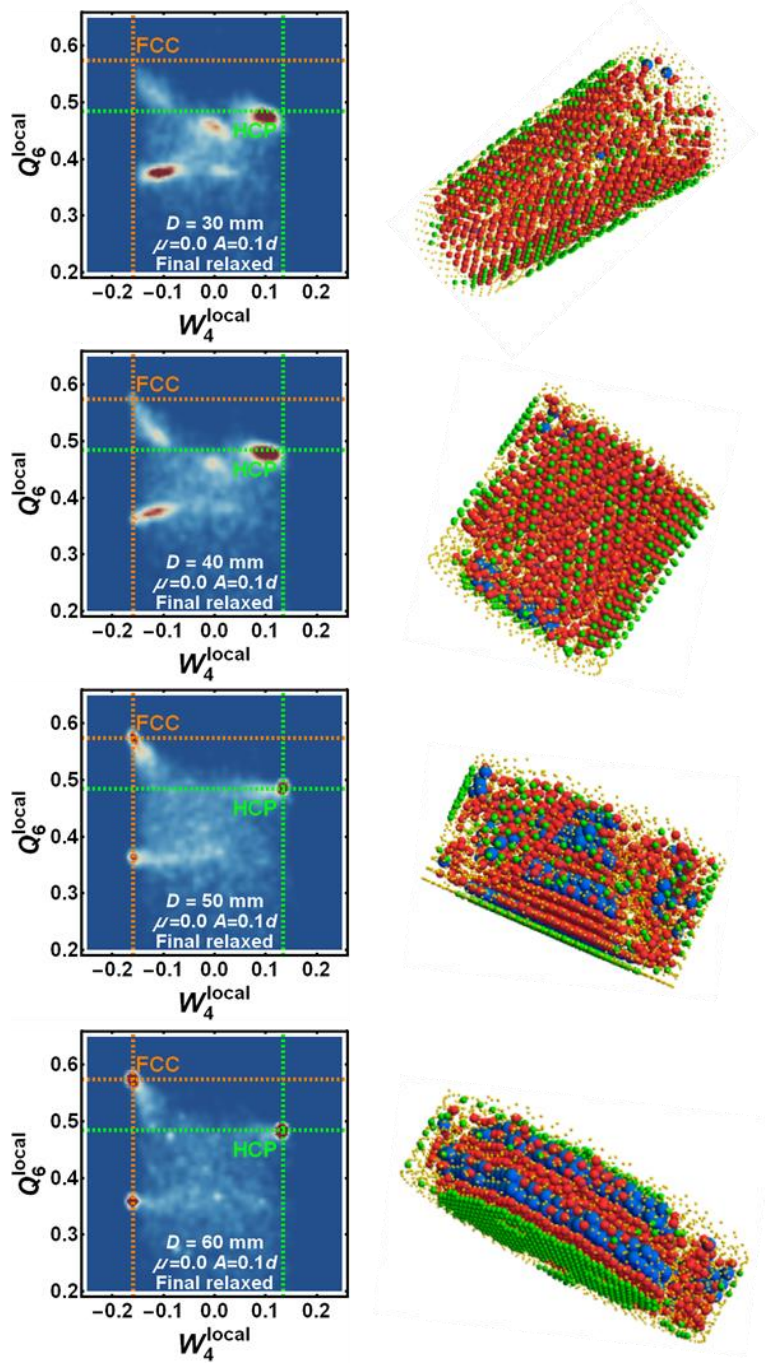

Figure 7. Density distributions of crystallised structures at the final relaxed state. The right column displays the corresponding packing structures with particles dyed according to the ($W_4^{local}$, $Q_6^{local}$) coordinates in red – HCP, blue – FCC, green – surface hexagon and yellow – others. The diameters of the particles are rescaled for visualisation purposes.

**Comparison with experimental data**

The effect of the boundary geometry on the granular crystallisation has been demonstrated in ideally frictionless media. It is necessary to clarify the influence of friction on crystallisation and compare it to available experimental data. Figure 8 shows for D30 the influence of friction on the evolution of the overall structural index $F_6$ during vibration. The initial state was kept identical for both the frictionless and frictional scenarios. The most obvious contrast between these two systems is the efficiency and degree of the granular crystallisation. The frictional energy losses between particles diminish the intensity of the vibration. Therefore, the crystallisation rate is lower in the frictional media and the required vibration time is longer to reach the final crystallised state. The simulated vibration duration (4000 periods for the longest vibration) is significantly shorter than the experimental study (40000 periods for the longest vibration), but the result still matches quantitatively.

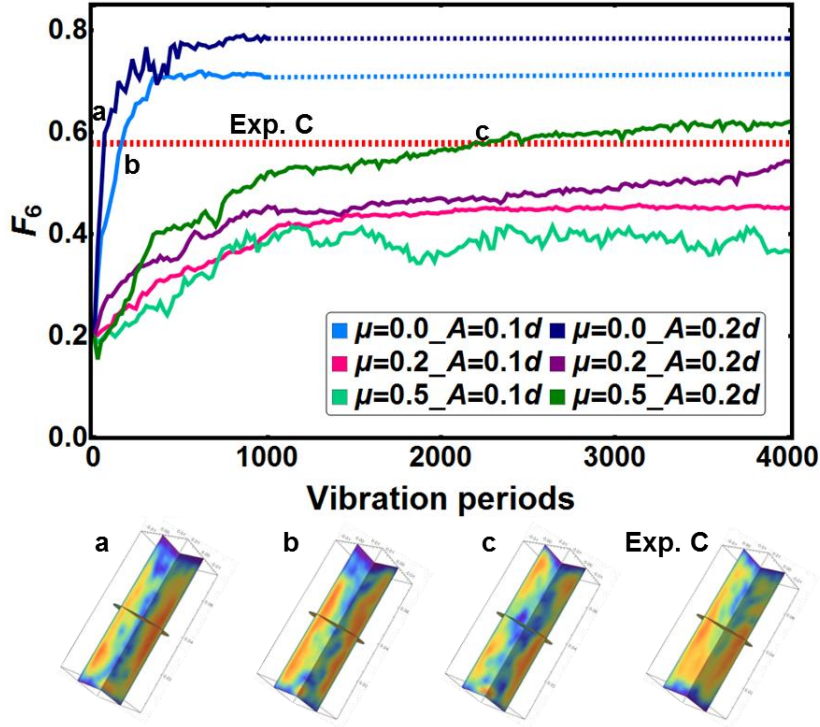

Figure 8. Top – Evolution of the structural index $F_6$ of the frictionless and frictional granular media. The top two dashed lines serve as the extension for the final states in the simulation and the middle one labelled with Exp. C represents the final state of the experiment performed with a vibration intensity $\Gamma = 2$ [39]. Bottom $-S_6$ spatial distributions of the labelled states in the simulated evolution and the experimental result. Friction coefficient ($\mu$) and amplitude ($A$) values for the simulations are displayed in the legend

Using the structural index obtained from the experimental results as a benchmark, the corresponding transient states in the simulation are extracted to reconstruct their temporary morphologies. Compared to the frictionless granular media, the frictional ones exhibit similar patterns of the $S_6$ spatial distributions, despite of the radial expansion, the crystalline regimes varying only slightly. Since the $S_6$ spatial distributions of different granular media match at the equivalent $F_6$, it is reasonable to argue that the structural evolution in the frictionless granular media represents a fairly complete crystallisation process for a given geometry. Hence, this morphological resemblance that demonstrates that friction scarcely influences the mechanisms of the granular crystallisation and merely hinders the crystallisation growth. Moreover, we have identified the inherent partially ordered regions existing at walls in various initial states conforming to recent XCT data (results not shown here). Nonetheless, such variation in the initial structures is of negligible influence for the wall effects and order development during vibration. Based on this argument, the experiments can be considered as an intermediate state in the evolution of the frictionless granular media. Due to the frictional force, the granular medium in the experiment is unable to reach the final state of the frictionless simulations, but maintains the same granular crystallisation mechanisms. The distribution obtained in the experiment shows crystalline patterns in the radial direction comparable with the theoretical predictions, while the results about the merging crystalline regime at the bottom of the container show a major difference between theory and experiment. This discrepancy is probably caused by the purely vertical vibration that ineffectively breaks the force chains sustaining the vertical perturbation in the simulations as well as the prolonged vibration in the experiments. The structural similarity between simulations and experiments is revealed by the $S_6$ distributions and the ($W_4^{local}$, $Q_6^{local}$) coordinate distributions as shown in Figure 9. All the extracted transient states and the experimental result follow the same shapes of the $S_6$ distributions with the final state of the frictionless granular media presenting the highest peak magnitude, which supports the previous argument. In addition, the simulated granular media resemble the ($W_4^{local}$, $Q_6^{local}$) distribution in the experiment. Similarly, the experimental medium forms two-dimensional hexagonal packing near the cylindrical wall and structures in a distorted hexagonal cubic fashion.

The second set of comparisons is made for the relatively flat granular media , designated as Exp. D in [39], performed with $\Gamma$ =2.8. Quantitative agreement is reflected by the different types of distributions used in this study. In addition, the experiment presents a similar and only one crystallisation mode demonstrated in the previous sections. With the elimination of the cylindrical wall crystallisation mode, the crystalline arrangement builds up from the bottom plane, and is characterised by the mixing of HCP and FCC planes with a few defects.

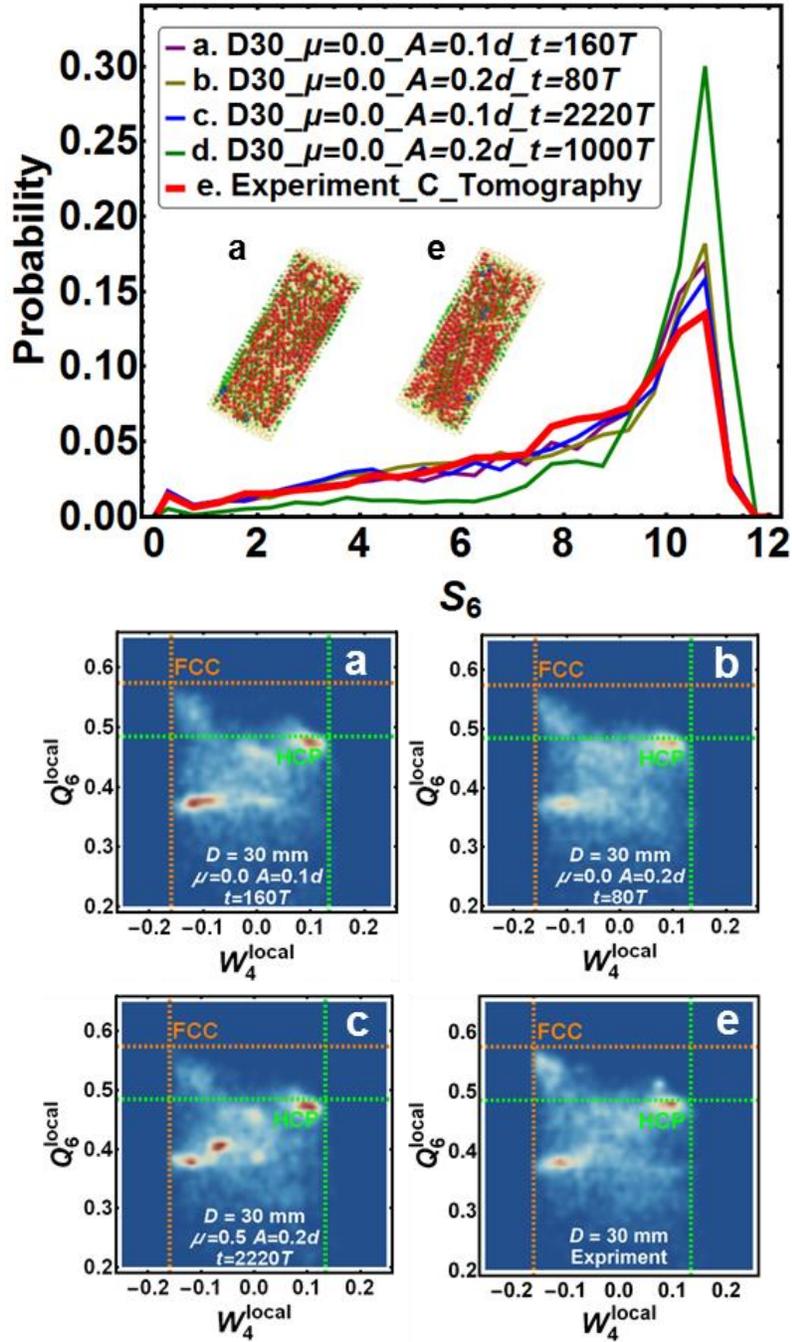

Figure 9. Top − $S_6$ distributions of the final state in the experiment Exp. C in [39] and the transient states in the simulations. Particles dyed according to the ($W_4^{local}$, $Q_6^{local}$) coordinates are displayed as insets in the $S_6$ distribution in red – HCP, blue – FCC and green – surface hexagon. Bottom – The corresponding ($W_4^{local}$, $Q_6^{local}$) coordinates distributions. Friction ($\mu$), amplitude ($A$) and duration ($t$) parameters for the simulations are displayed in the legend.

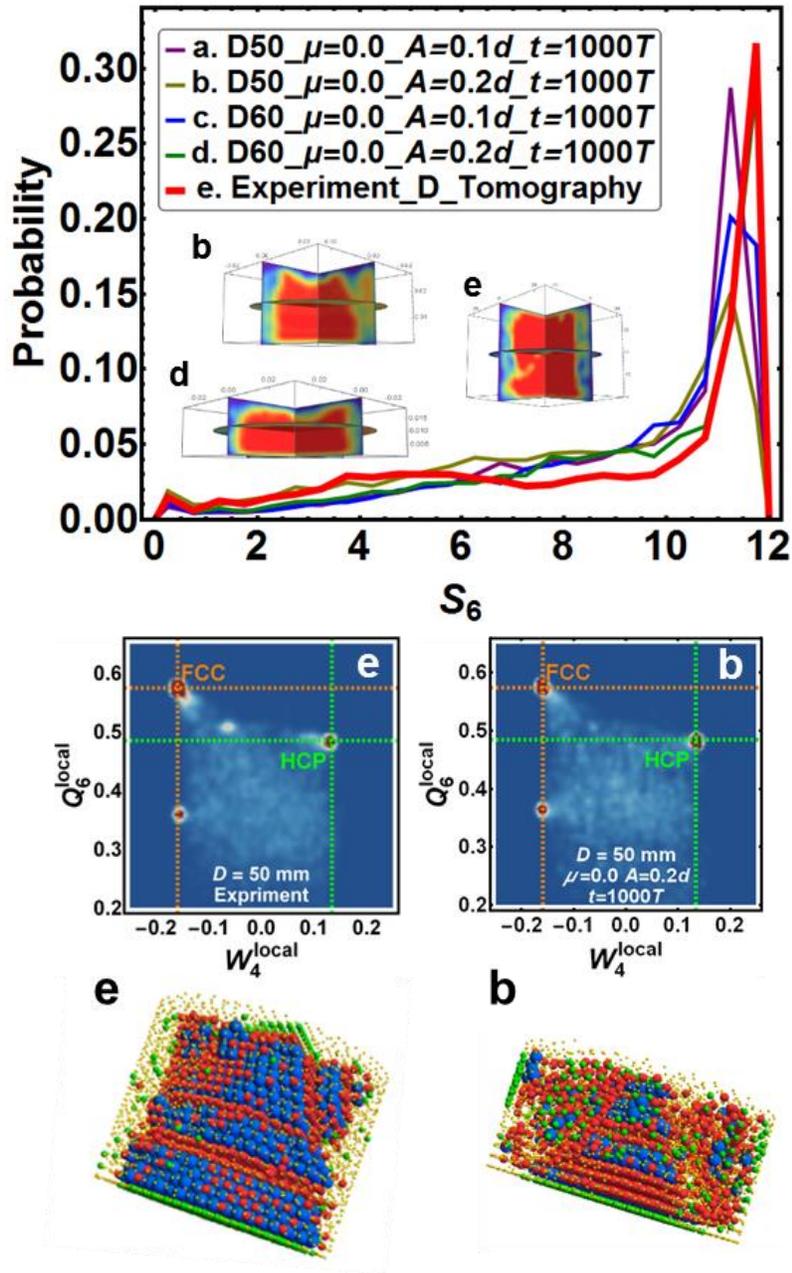

Figure 10. Top – $S_6$ distributions of the final state in the experiment Exp. D in [39] and the simulations along with the selected $S_6$ spatial distributions in the insets. Bottom – The corresponding ($W_4^{local}$, $Q_6^{local}$) coordinates distributions and the packing of dyed particles. Friction ($\mu$), amplitude ($A$) and duration ($t$) values for the simulations are displayed in the legend.

## Conclusion

Granular crystallisation has been investigated from the entire system scale down to the individual particle scale in confined granular matter. The results clearly show that vibration naturally brings about a disorder-to-order transition, proving that order formation during the transient evolution promotes granular crystallisation independent of densification.

The crystallisation process and the role of the wall effect are explained by coarse-graining approach. Internal nucleation growth is restrained due to the purely repulsive interactions, and crystallisation from walls is preferential. The wall effect produces two-dimensional hexagonal packing as a growth template but the following growth of crystalline arrays divides into a cylindrical mode and a bottom mode. In the cylindrical mode, the crystallised structure can be considered as a distorted HCP structure, while in the bottom mode, a mixture of finer HCP and FCC structure is identified. Depending on the geometry, $D/d$ and $H/D$, competition between these two crystallisation modes falls into three phases during vibration. By increasing $D$, the bottom mode crystallisation gradually dominates, leading the crystalline regime to penetrating throughout the entire granular media in the axial direction. In the other case, i.e., when $D$ decreases, the crystalline regions growing in the radial direction towards the axis are promoted from the cylindrical mode. Increasing the amplitude of vibration enhances the efficiency of the crystallisation, which can raise the competition level and leads to a sole crystallisation phase, which is commonly seen in experiments. Through the particle scale characterisation we conclude that particle are driven to lodge themselves in structures with as many contacts as possible, because such structures provide sufficient collisions to dissipate kinetic energy and maintain the stability of the granular packing. The relation between granular crystallisation and granular temperature is further explored and it is seen that granular crystallisation leads not only to structural but also to dynamic order.

In this study, granular crystallisation induced by vibration is proved to follow basic processes resulting in a predictable final structure. These results suggest that mechanical, thermal, electrical and other structure related effective properties can be modified by controlling vibration, motivating the continued study of granular crystallisation. Further research can be conducted to explore the physical mechanisms of inducing granular crystallisation. Meanwhile, this study shows the crystallisation can be connected to the statistic description of granular matter. Thus, it would be interesting to seek more precise correlations between granular crystallisation and granular statistics.